\documentclass[aps,groupaddress,superscriptaddress,reprint,amsmath,amssymb,graphicx]{revtex4-1}
\usepackage{graphicx}
\usepackage{physics}
\usepackage{bm}
\usepackage{bbold}
\usepackage{xcolor}
\begin{document}
\preprint{APS/123-QED}

\title{Mutual information disentangles interactions from changing environments}

\author{Giorgio Nicoletti}
\affiliation{Laboratory of Interdisciplinary Physics, Department of Physics and Astronomy ``G. Galilei'', University of Padova, Padova, Italy}
\author{Daniel Maria Busiello}
\affiliation{Institute of Physics, \'Ecole Polytechnique F\'ed\'erale de Lausanne - EPFL, 1015 Lausanne, Switzerland}

\begin{abstract}
\noindent Real-world systems are characterized by complex interactions of their internal degrees of freedom, while living in ever-changing environments whose net effect is to act as additional couplings. Here, we introduce a paradigmatic interacting model in a switching, but unobserved, environment. We show that the limiting properties of the mutual information of the system allow for a disentangling of these two sources of couplings. Further, our approach might stand as a general method to discriminate complex internal interactions from equally complex changing environments.
\end{abstract}

\maketitle
Real-world systems exhibit interactions among their internal degrees of freedom. Furthermore, they are usually coupled with a noisy, ever-changing environment. Modeling together these two distinct contributions is often a problem too hard to be tackled, and a common approach would prescribe to simply ignore environmental effects altogether.

In the last twenty years, however, it was realized in many different fields that the effects of a noisy environment are often as fundamental as the internal interactions. Biological systems \cite{Hilfinger2011, Tsimring2014}, biochemical \cite{dass2021} and gene regulatory networks \cite{Swain2002, Thomas2014, Bowsher2012}, swarming, oscillatory, and ecological systems \cite{Nosuke2004,Pimentel2008,Zhu2009} are only a few examples of noisy interacting systems living in an equally noisy environment, and being consequently affected by it. In the last years, it has been also shown that many observed properties believed to be distinctive of neural interactions can be solely explained by an environmental-like dynamics that affects all neurons in the same way \cite{Touboul2017, Ferrari2018, Nicoletti2020, Mariani2021}. Alternatively, some of these properties might be stochastic by nature and not reflect any particular feature of the underlying degrees of freedom \cite{Martinello2017}.

From a different perspective, crucial non-equilibrium features in chemical systems, such as thermophoresis \cite{piazza,liang2021}, and pattern formation \cite{turing}, have been recently shown to be sheer consequences of the interplay between environmental and internal interactions acting on different time-scales \cite{busiello2020}.

To make things more interesting, an ever-growing wealth of data is populating the realm of biological, chemical and neural systems, thus fueling the possibility of a direct extrapolation of some properties belonging to the underlying dynamics. In fact, when dealing with experimental data, it is not unusual to solve a given inverse problem, for example using a maximum entropy principle \cite{Schneidman2006, Mora2010, Bialek2012}, to reconstruct the interactions between the internal degrees of freedom that shape the observed behavior. However, one might ask whether these reconstructed effective couplings could possibly be a pure consequence of nothing but our ignorance about the unobserved environment in which the system lives. This question is often particularly hard to assess, as effective interactions arise even in non-interacting systems under the influence of a correlated noise \cite{Lise1999}.

In this work, we introduce a complete dynamical model, which includes both the internal dynamics, i.e. the one stemming from internal \textcolor{black}{physical} couplings, and stochastic environmental changes. While the internal dynamics is independent of the environment, and fully determined by system features, the environmental changes affect model parameters shared by all degrees of freedom. \textcolor{black}{Recently, this problem has gained momentum from a theoretical perspective in different contexts \cite{Bressloff2016, Hufton2016, Bressloff2017, Grebenkov2019}, but the general question of how we can possibly disentangle the effects of internal interactions from those of a stochastic environment is very much open and elusive.} Here, we consider the paradigmatic case of an environment affecting only the diffusion coefficient, hence changing the stochastic variability of the dynamics. \textcolor{black}{Nonetheless, our modeling approach can be immediately generalized to diverse scenarios, from more complicated environments to spatially inhomogeneous media \cite{Chechkin2017}.}

 We will directly tackle the disentangling problem by using the mutual information to quantify the similarity between different interacting processes. \textcolor{black}{In fact, the mutual information captures all cross-dependencies between two random variables \cite{ThomasCover2006}}. We will show that in the presence of linearized interactions the mutual information of the whole system does encode both internal and environmental couplings as distinct contributions, and that they can always be fully disentangled in suitable limits. Although characterizing the specific nature of internal interactions through mutual information remains a challenge, our results suggest that fast-varying environments might reveal the presence of underlying \textit{real} couplings in any general system.

The mutual information between two stationary processes $x_1(t)$ and $x_2(t)$ is \textcolor{black}{the Kullback-Leibler divergence between their joint stationary probability distribution $p(x_1, x_2)$ and the product of their marginalized stationary distributions $p(x_1)p(x_2)$,}
\begin{align}
\label{eqn:mutual}
    I & = \int dx_1dx_2 \, p(x_1, x_2) \log \frac{p(x_1, x_2)}{p(x_1) p(x_2)} \nonumber \\
    & = H_1 + H_2 - H_{12},
\end{align}
\textcolor{black}{where $H_{12}$ is the differential entropy of the joint distribution, and, similarly, $H_\mu$ is the differential entropy of the marginalized probability distributions for $\mu = 1, 2$.}

In order to fix the ideas, let us consider the paradigmatic example of two interacting Ornstein-Uhlenbeck processes \cite{gardiner2004}. This particular choice is twofold. First, an Ornstein-Uhlenbeck process is one of the simplest multidimensional stochastic process with a non-trivial stationary distribution. Second, Ornstein-Uhlenbeck processes can often be seen as a linearization of more complex, non-linear internal interactions. \textcolor{black}{Therefore, we introduce an internal dynamics by means of an interaction matrix $\vb{A}$ between the internal degrees of freedom $x_1(t)$ and $x_2(t)$. Then, we consider an archetypal description of the environmental changes, which we will regard as unobserved degrees of freedom acting on both $x_1$ and $x_2$ in the same way.} At all times, $x_1$ and $x_2$ share the same diffusion coefficient, and the diffusion coefficient itself is a stochastic variable. In particular, we take it to be a dichotomous process $D_{i(t)}$ between the states $i \in \{-,+\}$, so that the diffusion coefficient jumps between two states $D_-$ and $D_+ > D_-$, with transition rates $W(- \to +) = w_+$ and $W(+ \to -) = w_-$. \textcolor{black}{All in all, our model can be written as the set of Langevin equations}
\begin{align}
\label{eqn:langevin}
    \dv{x_\mu}{t} = -\sum_{\nu}A_{\mu\nu}\frac{x_\nu}{\tau} + \sqrt{2 D_{i(t)}} \xi_\mu(t)
\end{align}
\textcolor{black}{where $i(t)$ is a realization of the jump process between $\{-, +\}$ and $\xi_1$ and $\xi_2$ are independent white noises with zero mean. Here, the environment is encoded in the two distinct diffusion coefficients $D_i$, whereas the internal couplings stem from the off-diagonal elements of $\vb{A}$. Our goal is to understand whether these two distinct contributions to the dynamics can be disentangled, and, if so, under which conditions.}

\textcolor{black}{With this aim in mind, let us begin with the simple case $\vb{A} = \mathbb{1}$, i.e. $x_1$ and $x_2$ do not interact, so that the only contribution to the mutual information has to come from the environmental changes}. The system is described by a joint p.d.f. $p_i(\vb{x}, t)$ to have values $\vb{x} = (x_1, x_2)$ at time $t$ and to be in the environmental state $i \in \{-, +\}$. This probability is governed by the Fokker-Planck equation
\begin{align}
    \label{eqn:fokker_planck}
    \partial_t p_i(\vb{x}, t) = & \sum_{\mu = 1}^2\partial_\mu\left[\frac{x_\mu}{\tau}p_i(\vb{x}, t)\right] + D_i\sum_{\mu= 1}^2 \,\partial_\mu^2 \, p_i(\vb{x}, t) + \nonumber\\
    & + \sum_{j \ne i} \left[w_i  p_j(\vb{x}, t) - w_j  p_i(\vb{x}, t)\right].
\end{align}
This model corresponds, for instance, to a switching environment in a chemical \cite{dass2021,liang2021} or biological \cite{Bowsher2012,Tsimring2014} system, or to different regimes of neural activity \cite{Mariani2021,Touboul2017}. \textcolor{black}{Furthermore, being related to ``diffusing diffusivity'' processes, it can also describe spatially disordered or inhomogeneous environments \cite{Chechkin2017, Wang2020}.} An extension to $N$ different processes $(x_1, \dots, x_N)$ and $M$ environmental states $i_1, \dots, i_M$ is possible once we choose a multivariate generalization of the mutual information (see the Supplemental Materials (SM) \cite{supplemental_material} for details). 

Let us note beforehand that the mutual information, Eq.~\eqref{eqn:mutual}, can only depend on dimensionless quantities. The relevant dimensionless parameters of this model are: (i) $\tau w_\mathrm{sum}$, where $w_\mathrm{sum} = w_+ + w_-$, which governs the time-scale separation between the \textcolor{black}{internal degrees of freedom} and the jump process of the environmental states; (ii) $w_-/w_+$, which determines the relative persistence of the environmental states; (iii) $D_-/D_+$, which describes the separation between the environmental states. Importantly, the joint probability does not depend on these three parameters' combinations only. Hence, to find a general solution to Eq.~\eqref{eqn:fokker_planck} proves to be a particularly challenging task. Therefore, we resort to a time-scale separation that corresponds to the two limits in which the jumps are either much faster or much slower than the relaxation time of $x_1$ and $x_2$ (see the SM \cite{supplemental_material}).

\begin{figure*}[t]
    \centering
        \includegraphics[width=0.9\textwidth]{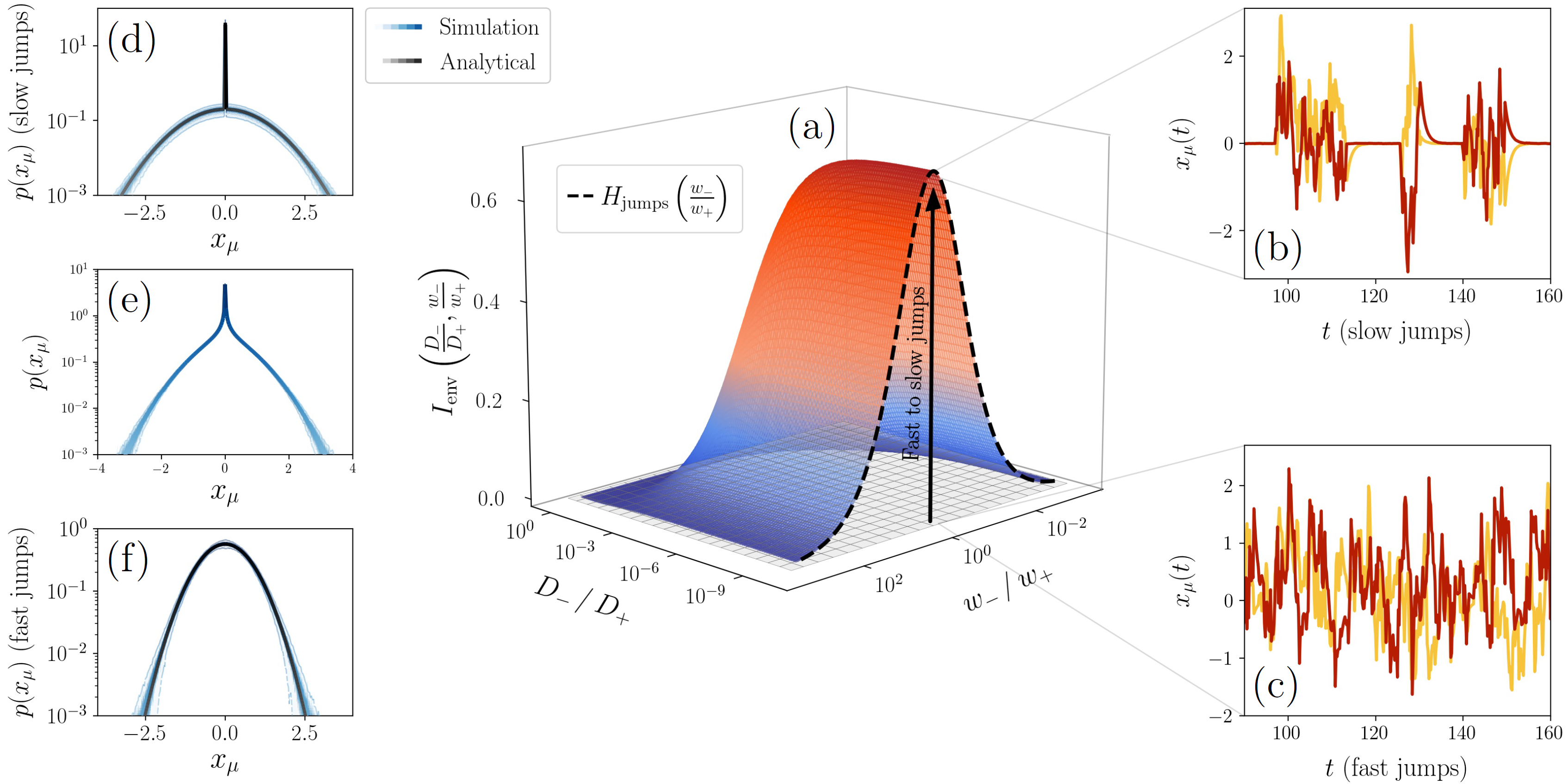}
    \caption{The environmental contribution to the mutual information as a function of $D_-/D_+$ and $w_-/w_+$. (a) The colored surface is the result of a Monte Carlo integration with importance sampling of the mutual information in the slow-jumps limit with $\tau w_{\rm sum}= 10^{-3}$(see Eq.~\eqref{eqn:mutual} and Eq.~\eqref{eqn:fast_jumps}). In the $D_-/D_+ \to 0$ limit, $I_\mathrm{env}$ becomes exactly $H_\mathrm{jumps}$, the black dashed line, which is also its maximum value. The gray plane is instead the mutual information in the fast-jumps limit, which always vanishes. (b) A realization of $x_1(t)$ and $x_2(t)$ (red and yellow curves) in the slow-jumps limit, at $w_-/w_+ = 1$ and $D_-/D_+ = 10^{-10}$. A bursty, coordinated behavior emerges due to the environmental changes. (c) Same, but in the fast-jumps limit, where both variables show a Brownian-like behavior. (d-f) Comparison between the marginalized probability distribution $p(x_\mu)$ from the simulated Langevin dynamics and the analytical distribution ((d) and (f) cases) for $D_- = 10^{-5}$, $D_+ = 1$, $\tau = 1$ in (d) the slow-jumps limit at $w_- = w_+ = 5 \cdot 10^{-4}$,  (f) the fast-jumps limit at $w_- = w_+ = 50$ and (e) in between at $w_- = w_+ = 0.5$.}
    \label{fig:mutual_extrinsic}
\end{figure*}

In a fast environment, we have $\tau w_\mathrm{sum} \gg 1$, and we find the stationary probability distribution
\begin{align}
\label{eqn:fast_jumps}
    p^F(x_1, x_2) & = \frac{1}{2\pi \tau\ev{D}_\pi} \exp[-\frac{x_1^2 + x_2^2}{2\tau \ev{D}_\pi}] \equiv p^F(x_1)p^F(x_2)
\end{align}
where $\ev{D}_\pi = (D_+w_+ + D_-w_-)/w_\mathrm{sum}$ plays the role of an effective diffusion coefficient, and the superscript $F$ refers to the fast-jumps regime. \textcolor{black}{Loosely speaking, this limit describes environmental changes affecting the internal degrees of freedom only on average, leaving the two processes independent. Hence, the joint probability factorizes and no mutual information arises (Figure \ref{fig:mutual_extrinsic})}.

The picture is {\color{black}markedly} different in the slow-jumps limit, when $\tau w_\mathrm{sum} \ll 1$. The stationary probability distribution is now given by the Gaussian mixture
\begin{align}
\label{eqn:slow_jumps}
    p^S(x_1, x_2) & = \frac{1}{2\pi \tau}\sum_{i}\left[\frac{\pi_i^s}{D_i} e^{-\frac{1}{2D_i\tau}\left(x_1^2 + x_2^2\right)} \right] \nonumber\\
    & = \pi_- p^S_-(x_1, x_2) + \pi_+ p^S_+(x_1, x_2)
\end{align}
where $\pi_i = w_i/w_\mathrm{sum}$ are the stationary probabilities of the jump process, and the superscript $S$ denotes the slow-jumps regime. It is clear that in this limit the two processes are not always independent. An example of a realization and its corresponding probability distribution is shown in Figure \ref{fig:mutual_extrinsic}b and \ref{fig:mutual_extrinsic}d, respectively. In the intermediate regime between the fast- and slow-jumps limits we cannot solve the Fokker-Planck equation explicitly, but a direct simulation of the Langevin equations \cite{Gillespie1996}, Eq.~\eqref{eqn:langevin}, shows that the resulting probability interpolates between Eq.~\eqref{eqn:slow_jumps} and Eq.~\eqref{eqn:fast_jumps} in a smooth fashion (Figure \ref{fig:mutual_extrinsic}d-f). Therefore, we will now focus on the slow-jumps limit, where we can tackle the problem analytically, and the mutual information takes non-zero values.

Even though no closed form exists for the entropy of a Gaussian mixture, from the bounds proposed in \cite{Kolchinsky2017} we can build the corresponding bounds on the mutual information starting from the Chernoff-$\alpha$ divergence and on the Kullback-Leibler divergence between the mixture components{\color{black}, reported in the SM} \cite{supplemental_material}. \textcolor{black}{Notably, both the upper bound $I_\text{env}^{S,\rm{up}}\left(D_-/D_+, w_-/w_+ \right)$ and the lower bound $I_\text{env}^{S,\rm{low}}\left(D_-/D_+, w_-/w_+ \right)$ on the mixture distribution $p^S(x_1, x_2)$ only depend on the dimensionless parameters $D_-/D_+$ and $w_-/w_+$. Moreover, although in general these bounds are not tight, they do saturate} in the limits $D_-/D_+ \to 0$ or $D_-/D_+ \to 1$ - and these limits are particularly significant. The former corresponds to drastic environmental changes, which lead to markedly different dynamics \textcolor{black}{and give rise to a bursty, seemingly coordinated behavior of the internal degrees of freedom. The latter, on the other hand,} describes the trivial case in which $D_-$ and $D_+$ are very similar and thus {\color{black}environmental} changes are effectively negligible. We end up with (see SM \cite{supplemental_material})
\begin{align}
\label{eqn:mutual_extrinsic_limits}
    I^S_\text{env}\left(\frac{w_-}{w_+}\right) =
    \begin{cases}
        -\pi_+ \log \pi_+ -\pi_- \log \pi_- &\, \text{if} \quad D_+ \gg D_- \\
        \,0 &\, \text{if} \quad D_+ \approx D_-
    \end{cases}
\end{align}
which, since the bounds saturate, are the exact limits of the mutual information in the slow-jumps regime. Clearly, when $D_-/D_+ \to 1$, the dynamics is insensitive to the environment, thus $x_1(t)$ and $x_2(t)$ are independent processes. \textcolor{black}{Instead, and interestingly}, the first line is nothing but the Shannon entropy of the jump distribution, $H_\mathrm{jumps}(w_-/w_+)$. A Monte Carlo integration of Eq. (\ref{eqn:mutual}) shows that $H_\mathrm{jumps}$ is also the maximum value of the mutual information that emerges due to the environment, see Figure \ref{fig:mutual_extrinsic}a. This result has a quite clear intuitive interpretation. \textcolor{black}{In fact, from an information-theoretic point of view, $H_\mathrm{jumps}$ quantifies precisely the information lost once we integrate out the stochastic environment, i.e. our ignorance about the system as a whole.}

So far, we have only considered the presence of an effective coupling emerging from environmental changes. Although our results have been derived for Ornstein-Uhlenbeck processes, they equivalently hold for the more general stochastic dynamics $\dot{x}_\mu(t) = f_\mu(x_\mu) + \sqrt{2 D_{i(t)}} \xi_\mu(t)$, even when $f_\nu \neq f_\mu$, \textcolor{black}{as we show in the SM \cite{supplemental_material}}. \textcolor{black}{Now, it is time to introduce back interactions between $x_1(t)$ and $x_2(t)$ by considering the case in which the matrix $\vb{A}$ in Eq.~\eqref{eqn:langevin} has non-zero off-diagonal entries. We will show that their contribution to the total mutual information, $I_{\rm tot}$, can be effectively disentangled from $I_{\rm env}$ under suitable limits.}

Let us consider the matrix
\begin{equation*}
    \vb A = \begin{pmatrix}
                1 & - g_1 \\
                - g_2 & 1
            \end{pmatrix}
\end{equation*}
\textcolor{black}{and assume that its eigenvalues have positive real parts, so that a stationary state exists \cite{Note1}}. Let us also assume, for the time being, that we are in the slow-jumps limit, so that we can solve the Langevin equations separately for $D_+$ and $D_-$ and then average them over $\pi_\pm$ as in Eq.~\eqref{eqn:slow_jumps}. The two solutions are multivariate Gaussian distributions, each one with a given covariance matrix $\bm{\Sigma}_\pm$.

\begin{figure}[t]
    \centering
    \includegraphics[width=0.45\textwidth]{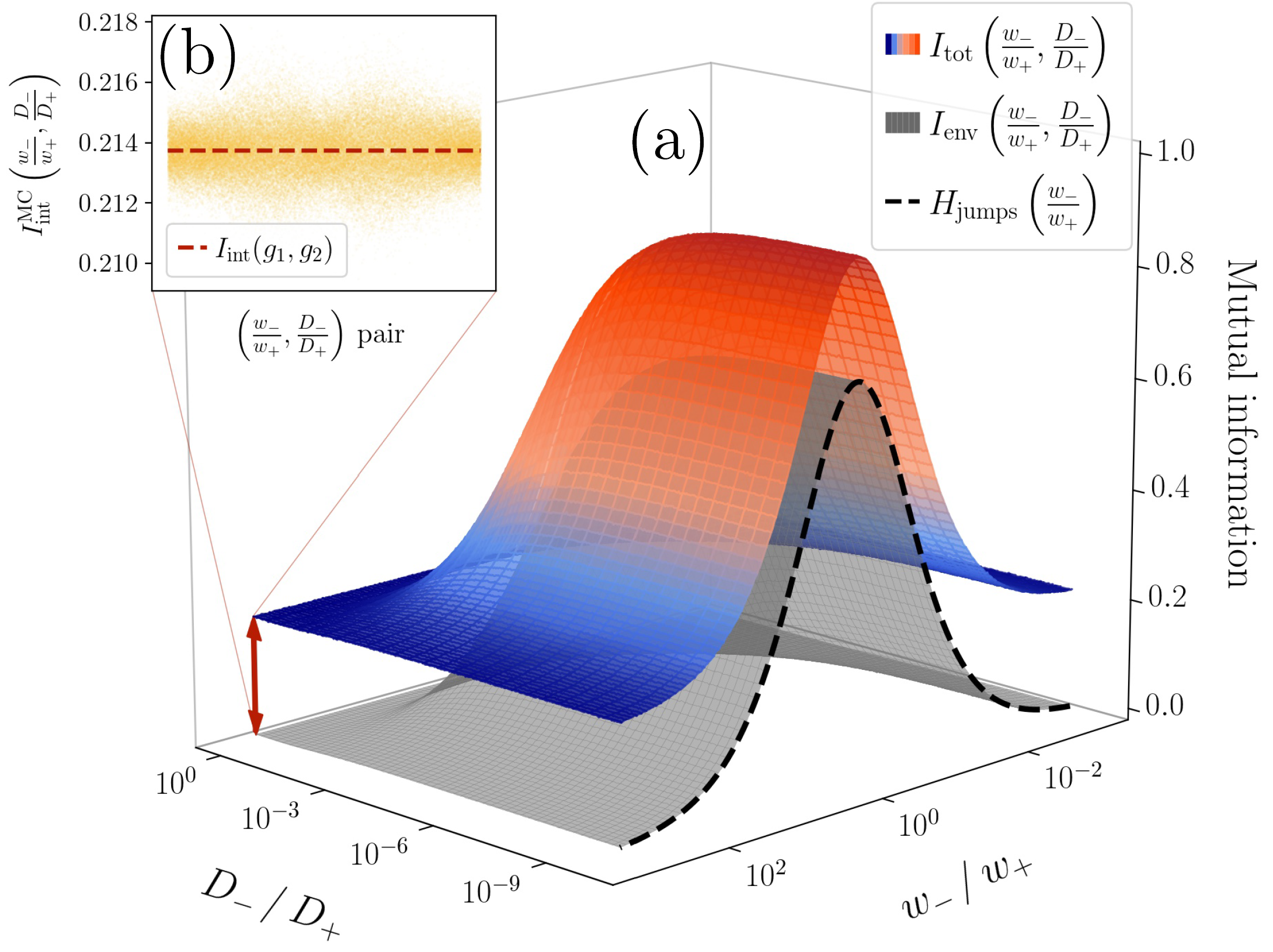}
    \caption{The total mutual information as a function of $D_-/D_+$ and $w_-/w_+$ at fixed $\tau w_{\rm sum}= 10^{-3}$, i.e. in the slow-jumps limit. (a) The colored surface is the result of a Monte Carlo integration with importance sampling of $I_\text{tot}$, in the slow-jumps limit, for the interacting model with $g_1 = 5 \tau$ and $g_2 = -0.1 \tau$. The gray surface is instead the non-interacting case, $I_{\rm env}$. The two contributions to the mutual information disentangle and the interactions simply result in a constant shift. (b) A comparison between the predicted shift $I_\text{int}(g_1, g_2)$, Eq.~\eqref{Iintg1g2}, and the difference of the Monte Carlo estimates of the two surfaces for every sampling point $(w_-/w_+, D_-/D_+)$, namely $I_\text{int}^{MC}(w_-/w_+, D_-/D_+)$.}
    \label{fig:mutual_intrinsic}
\end{figure}

In order to try and disentangle the environmental contribution, which is due to $D_i$, and the one stemming from internal interactions, due to the off-diagonal elements of $\vb A$, we write the covariance matrices as $\bm\Sigma_i = D_i \tilde{\bm\Sigma}$. The matrix $\tilde{\bm\Sigma}$ then solves the Lyapunov equation (see the SM \cite{supplemental_material})
\begin{equation}
    \frac{1}{2}\left[\vb{A}\tilde{\bm\Sigma} + \tilde{\bm\Sigma} \vb{A}^T\right] = \mathbb{1}
\end{equation}
which only depends on interactions and not on the jump dynamics, nor on $D_i$. We are now able to bound the mutual information as
\begin{align}
\label{eqn:up/low}
    I^{S,\rm{up/low}}_{\rm tot} = I_\text{int}\left(\{g_\mu\}\right) + I^{S,\rm{up/low}}_\text{env}\left(\frac{D_-}{D_+}, \frac{w_-}{w_+}\right)
\end{align}
where
\begin{align}
   I_\text{int}\left(\{g_\mu\}\right) = \frac{1}{2} \log\left[1 - \frac{4}{4 + (g_1 - g_2)^2}+ \frac{1}{1 - g_1 g_2} \right]
   \label{Iintg1g2}
\end{align}
is the contribution to the mutual information due to the internal interactions only, as we show in the SM \cite{supplemental_material}. \textcolor{black}{Notably, $I_\text{int}$ is also the sole contribution in the fast-jumps limit, since no environmental contribution is present to begin with.} On the other hand, in the slow-jumps limit we can write the two limits, as in Eq.~\eqref{eqn:mutual_extrinsic_limits},
\begin{align}
\label{eqn:mutual_tot_limits}
    I^S_\text{tot}\left(\{g_\mu\}, \frac{w_-}{w_+}\right) =
    \begin{cases}
        H_\mathrm{jumps} + I_\text{int}\left(\{g_\mu\}\right) & \text{if} \, D_+ \gg D_- \\
        I_\text{int}\left(\{g_\mu\}\right) & \text{if} \, D_+ \approx D_-
    \end{cases}
\end{align}
where the environmental bounds saturate. Finally, in the intermediate regime between the fast- and the slow-jumps limits, a Monte Carlo integration of Eq.~\eqref{eqn:mutual} shows that the presence of linear internal interactions does simply shift the mutual information with respect to the non-interacting case (Figure \ref{fig:mutual_intrinsic}).

Therefore, our results suggest that $I_\mathrm{tot}$ receives two distinct contributions - one from the environment, $I_\mathrm{env}$, and one from the internal linearized interactions, $I_\mathrm{int}$ - disentangled in form:
\begin{equation}
    I_\text{tot}\left(\{g_\mu\},\frac{D_-}{D_+}, \frac{w_-}{w_+}\right) = I_\text{int}(\left\{g_\mu\}\right) + I_\text{env}\left(\frac{D_-}{D_+}, \frac{w_-}{w_+}\right).
\end{equation}
Although the equation above holds analytically in the fast-jumps regime, and where the slow-jumps bounds Eqs.~\eqref{eqn:up/low}-\eqref{eqn:mutual_tot_limits} saturate, its validity has been numerically shown in the entire range of parameters. Furthermore, even if the interactions are non-linear, we show in the SM \cite{supplemental_material} that in the fast-jumps limit the environmental contribution vanishes exactly. \textcolor{black}{Hence, and independently of the underlying interactions, any non-zero mutual information in the fast-jumps limit acts as a fingerprint of the presence of internal couplings.}

This result is extremely interesting. In fact, although the environmental states, identified by $D_-$ and $D_+$ in our model, are usually not experimentally accessible, it might be possible to characterize the frequency of the environmental changes. Neural activity originated by external stimuli \cite{neuron1, neuron2, neuron3}, stirring in chemical conglomerates \cite{Viedma2011}, temperature-activated chemical reactions in solutions \cite{astumian,dass2021}, and population growth \cite{frey,kussell,microbial}, are only a few examples in which our framework might apply. Even if fast-varying environments have been shown to be informative, our approach might provide hints about the presence of interactions even away from the fast-jumps limit, by bounding the environmental contribution to the mutual information. This intriguing perspective will be investigated in future works.

\textcolor{black}{Although we focused on a paradigmatic, but rather comprehensive, physical model, let us note that these ideas have a much larger scope, and that disentangling the different dependencies of a system is a far-reaching question. Techniques such as Bayesian networks and other probabilistic graphical models have been successfully used in biological data, for instance to disentangle different sources of interactions and dependencies in general \cite{Burger2008, Burger2010, layeghifard2017}. Connections may be also drawn to machine learning and artificial neural networks, particularly in the context of learning disentangled representation of the data, i.e. representations in which the informative latent factors are described by a factorized distribution \cite{kim2018, Chen2018, Locatello2019, Raban2020}, or in generative models with latent variables, such as switching state space models \cite{fox2009, linderman2017}. The environment in our model, in fact, can be seen as a latent variable, i.e. unobserved and independent on the observed degrees of freedom, while affecting the observed dynamics. Unlike the one presented here, these approaches are often harder to interpret and are less prone to the derivation of exact results, even though they remain extremely powerful in dealing with experimental data. Hence, a possible future perspective is to combine the physical interpretability and the analytical procedures behind our work together with tools from machine learning and data-driven approaches. This could lead to promising results in the quest of meaningfully disentangle the different sources of dependencies that emerge in complex systems.}

Furthermore, there are several more possible extensions to this study. One might ask whether a stochastic environment can be mapped into a set of effective couplings with defined properties, and if such couplings can be distinguished from the internal ones. Additionally, an important \textcolor{black}{and immediate} generalization of our framework is to allow the environment to be a continuous variable. These problems, in principle, can be treated from a field-theoretical perspective, where the marginalization over the environment gives rise to new interaction vertexes that are not present in the original theory, i.e. before the marginalization. Ideally, this could allow for a much more general framework amenable to analytical treatments.

On the other side, the ability to analytically deal with a class of stochastic processes with tools of information theory, as shown here, opens up many fascinating possibilities. A particularly appealing question is what happens when, instead of considering a stochastic environment, the system undergoes an external perturbation - notably, how the latter changes the information content and how such information evolves over time. A first step towards this direction might be to consider two diffusion processes in a finite domain that undergo a single stochastic jump, and to study the persistence of the mutual information as a function of time, domain size and boundary conditions of the system.

Ultimately, we believe that this work draws a path towards a deeper understanding of the different sources of couplings in real-world systems. Indeed, it is a starting point to elucidate the relations between their internal complexity and possibly equally complex, but unobserved, ever-changing environments.

\begin{acknowledgments}
\noindent We acknowledge A. Maritan, S. Azaele, and S. Suweis for insightful discussions, valuable suggestions, and a careful reading of the manuscript.
\end{acknowledgments}

\pagebreak

\newpage
\widetext
\pagebreak

\setcounter{equation}{0}
\setcounter{figure}{0}
\setcounter{table}{0}
\setcounter{page}{1}
\setcounter{section}{0}
\setcounter{subsection}{0}
\makeatletter
\renewcommand{\theequation}{S\arabic{equation}}
\renewcommand{\thefigure}{S\arabic{figure}}
\renewcommand{\thesection}{S\Roman{section}} 


\begin{center}\Large{Supplemental Material: ``Mutual information disentangles interactions from changing environments'' }\end{center}

\section*{A. Generalization of the environmental dynamics to $N$ processes and $M$ jumps}
Let us consider a set of $N$ Langevin equations of the form
\begin{equation}
    \label{eqn:langevin_equations}
    \dv{x_\mu}{t} = - f_\mu(x_\mu) + \sqrt{2D_{i(t)}} \, \xi_\mu(t) \quad\quad \mu = 1 \dots, N
\end{equation}
where $\xi_\mu(t)$ are independent white noises such that $\ev{\xi_\mu(t)} = 0$ and $\ev{\xi_\mu(t_1)}\xi_\nu(t_2) = \delta_{\mu\nu} \delta(t_1 - t_2)$. As in the main text, all the variables $\vb{x}$ share the same diffusion coefficient $D_{i(t)}$, where $i(t)$ is a discrete stochastic process with $M$ states. The probability $\Pi_i(t) = \mathbb{P}[i(t) = i]$ is described by the master equation
\begin{equation}
    \label{eqn:jump_process}
    \partial_t \Pi_i(t) = \sum_{j = 1}^M \left[W(j \to i) \pi_j(t) - W(i \to j) \pi_i(t)\right]
\end{equation}
that is independent on all $x_\mu$.

We can write the corresponding Fokker-Planck equations as
\begin{align}
    \label{eqn:fokker_planck}
    \partial_t p_i(\vb{x}, t) = & \sum_{\mu = 1}^N\partial_\mu\left[f_\mu(x_\mu)p_i(\vb{x}, t)\right] + D_i\sum_{\mu= 1}^N \,\partial_\mu^2 \, p_i(\vb{x}, t) + \sum_{j = 1}^M \left[W(j \to i)  p_j(\vb{x}, t) - W(i \to j)  p_i(\vb{x}, t)\right]
\end{align}
where we used the shorthand notation $\partial_{x_\mu} := \partial_\mu$ and $\vb{x} = (x_1, \dots, x_N)$. These are $M$ equations for each of the discrete states of $D_i$. Similarly to the main text, the first row is associated to the continuous stochastic process described by the Langevin equations Eq. (\ref{eqn:langevin_equations} at a given diffusion coefficient, whereas the second row describes the jump process of Eq. (\ref{eqn:jump_process}) for the diffusion coefficient itself.

Since we have $N$ variables, we need to choose a suitable generalization of the mutual information. Such generalizations, however, are troublesome from an information-theoretic perspective \cite{ThomasCover2006}. Since the variables are not interacting but they only share the same environmental diffusion coefficient, we shall be interested in the factorizability of the joint probability distribution $p(\vb{x})$ with respect to its full factorization, that is
\begin{equation}
\label{eqn:multivariate_information}
    I_N = \int \prod_{\mu = 1}^N dx_\mu p(x_1, \dots, x_N) \log\frac{p(x_1, \dots, x_N)}{\prod_{\mu = 1}^N p(x_\mu)} = \sum_{\mu = 1}^N H_\mu - H_{1, \dots, N}
\end{equation}
where $H_\mu$ and $H_{1, \dots, N}$ are the entropies of the corresponding probability distributions. This is nothing but the Kullback-Leibler divergence between the joint probability distribution and the product of the single-variable distributions. Thus, this quantity is always positive and for $N=2$ it gives exactly the mutual information.

\section*{B. Fast- and slow- jumps limit for $N$ processes and $M$ jumps}
We assume that the Langevin equations are associated with a timescale $\tau$ - e.g. the fastest timescale of $\vb{x}$ - whereas the jump process happens at a timescale $\tau_{\rm jumps}$ - e.g. $\tau_{\rm jumps} = \left(\sum_{i \ne j} W(i\to j)\right)^{-1}$. Hence we can write the rescaled equation
\begin{align}
    \label{eqn:fokker_planck_rescaled}
    \partial_t p_i(\vb{x}, t) = & \frac{1}{\tau} \sum_{\mu = 1}^N \biggl[\partial_\mu\left[\tilde{f}_\mu(x_\mu)p_i(\vb{x}, t)\right] + \tilde{D}_i \,\partial_\mu^2 \, p_i(\vb{x}, t) \biggl]+ \nonumber \\
    & + \frac{1}{\tau_{\rm jumps}}\sum_{j = 1}^M \left[\tilde W(j \to i)  p_j(\vb{x}, t) - \tilde W(i \to j)  p_i(\vb{x}, t)\right]
\end{align}
where $\tilde f_\mu := \tau f_\mu$, $\tilde D_i := \tau D_i$ and $\tilde W(i \to j) := \tau_{\rm jumps} W(i \to j)$.

Let us now assume that the jump process is faster, i.e. $\tau_{\rm jumps}/\tau = \gamma \ll 1$. In this scenario, it makes sense to rescale the slow timescale $\tau$, namely $t \to t/\tau$ and to look for a solution of the form
\begin{equation}
    \label{eqn:expansion_fast_jumps}
    p_i(\vb{x}, t) = p_i^{(0)}(\vb{x}, t) + \gamma \, p_i^{(1)}(\vb{x}, t) + \mathcal{O}(\gamma^2).
\end{equation}
Thus, we end up with the Fokker-Planck equation
\begin{align}
    \partial_t p_i^{(0)} = & \sum_{\mu = 1}^N \biggl[\partial_\mu\left[\tilde{f}_\mu(x_\mu)p_i^{(0)}\right] + \tilde{D}_i \,\partial_\mu^2 \, p_i^{(0)} \biggl]+ \nonumber\\ 
    & + \sum_{j = 1}^M \left[\tilde W(j \to i)  p_j^{(1)} - \tilde W(i \to j) p_i^{(1)}\right] + \nonumber\\
    & + \frac{1}{\gamma}\sum_{j = 1}^M \left[\tilde W(j \to i)  p_j^{(0)} - \tilde W(i \to j) p_i^{(0)}\right] + \mathcal{O}(\gamma).
\end{align}
At the leading order we simply have
\begin{equation*}
    0 = \sum_{j = 1}^M \left[\tilde W(j \to i)  p_j^{(0)}(\vb{x}, t) - \tilde W(i \to j) p_i^{(0)}(\vb{x}, t)\right]
\end{equation*}
which is the stationary condition of the jump process. Hence we can assume that the zero-th order solution can be factorized as $p_i^{(0)}(\vb{x}, t) = \pi_i P(\vb{x}, t)$, where
\begin{equation*}
    0 = \sum_{j = 1}^M \left[\tilde W(j \to i)  \pi_j^s - \tilde W(i \to j) \pi_i\right]
\end{equation*}
defines the dependence on the $i$-th index and we only need to find $P(\vb{x}, t)$.

At the order $\mathcal{O}(1)$ we can sum over $i$ to find
\begin{align*}
    \partial_t P(\vb{x}, t) = & \sum_{\mu = 1}^N \biggl[\partial_\mu\left[\tilde{f}_\mu(x_\mu)P(\vb{x}, t)\right] + \left(\sum_i\pi_i\tilde{D}_i\right) \,\partial_\mu^2 \, P(\vb{x}, t) \biggl]
\end{align*}
which gives us the solution for $P(\vb{x}, t)$ as the solution for the Langevin equations in Eq. (\ref{eqn:langevin_equations}) with an effective diffusion coefficient $\sum_i\pi_i\tilde{D}_i$. This is nothing but a set of $N$ independent equations that can be solved separately. Thus we find that the solution in the limit of fast jumps is simply given by the factorization
\begin{align}
    \label{eqn:solution_fast_jumps}
    p(\vb{x}, t) = \sum_i p_i^{(0)}(\vb{x}, t) = \prod_{\mu = 1}^N g_\mu(x_\mu, t)
\end{align}
where $g_\mu(x_\mu, t)$ solves the one-dimensional equation
\begin{align*}
    \partial_t g_\mu(x, t) = \partial_x[\tilde f_\mu(x)g_\mu(x, t)] + \left(\sum_i\pi_i\tilde{D}_i\right) \partial_x^2g_\mu(x, t).
\end{align*}
We immediately see that in this limit all the variables $x_\mu$ are independent and Eq. (\ref{eqn:multivariate_information}) is zero.

We are also interested in the opposite limit, where the Langevin equations in Eq. (\ref{eqn:langevin_equations}) relax faster to their own stationary state, that is in the limit $\tau/\tau_{\rm jumps} := \delta \ll 1$. As before, we rescale $t \to t/\tau_{\rm jumps}$ and we end up with the Fokker-Planck equation
\begin{align}
    \partial_t p_i^{(0)} = & \frac{1}{\delta}\sum_{\mu = 1}^N \biggl[\partial_\mu\left[\tilde{f}_\mu(x_\mu)p_i^{(0)}\right] + \tilde{D}_i \,\partial_\mu^2 \, p_i^{(0)} \biggl]+ \nonumber\\ 
    & + \sum_{\mu = 1}^N \biggl[\partial_\mu\left[\tilde{f}_\mu(x_\mu)p_i^{(1)}\right] + \tilde{D}_i \,\partial_\mu^2 \, p_i^{(1)} \biggl]+ \nonumber\\ 
    & + \sum_{j = 1}^M \left[\tilde W(j \to i)  p_j^{(0)} - \tilde W(i \to j) p_i^{(0)}\right] + \mathcal{O}(\delta).
\end{align}
Once more the leading order carries no temporal dependence so we assume that $p_i^{(0)}(\vb{x}, t) = \pi_i(t)P_i^s(\vb{x})$. Furthermore, we can write $P_i^s(\vb{x}) = \prod_\mu P_{\mu}^s(x_\mu, D_i)$
 where $P_{\mu}^s(x, D_i)$ solves
\begin{align*}
    0 = \partial_x\left[\tilde{f}_\mu(x)P_{\mu}^s(x, D_i)\right] + \tilde{D}_i \,\partial_x^2 \,P_{\mu}^s(x, D_i)
\end{align*}
which are nothing but the (independent) stationary solutions of each of the Langevin equations in Eq. (\ref{eqn:langevin_equations}) at constant diffusion coefficient.

The $\mathcal{O}(1)$ order, after an integration over $\vb{x}$, gives instead
\begin{align*}
    \partial_t \pi_i(t) = \sum_{j = 1}^M \left[\tilde W(j \to i) \pi_j(t) - \tilde W(i \to j) \pi_i(t)\right]
\end{align*}
which leads to the overall solution
\begin{align}
    \label{eqn:solution_slow_jumps}
    p(\vb{x}, t) = \sum_{i = 1}^M \left[\pi_i(t) \prod_{\mu = 1}^N P_{\mu i}^s(x_\mu)\right]
\end{align}
where we denote $P_{\mu}^s(x_\mu, D_i)$ with $P_{\mu i}^s(x_\mu)$ for the sake of brevity. In this limit, the variables $x_\mu$ are not independent anymore and indeed their joint distribution is a mixture distribution.

\section*{C. Multivariate information for $N$ processes and $M$ jumps}
We now consider the limit of slow jumps, so that we end up with a probability distribution that is not trivially factorizable. Let us write the stationary limit of the one variable probability distributions as
\begin{equation}
\label{eqn:mixture_1D}
    p_\mu(x) = \sum_{i=1}^M \pi_i P_{\mu i}^s(x)
\end{equation}
and the $N$ variables probability distribution as
\begin{equation}
\label{eqn:mixture_2D}
    p_{1, \dots, N}(\vb{x}) = \sum_{i=1}^M \pi_i \prod_{\mu = 1}^N P_{\mu i}^s(x_\mu).
\end{equation}
In order to study the multivariate information in Eq. (\ref{eqn:multivariate_information}) we need to bound the entropies of these distributions, which do not admit a closed form. From \cite{Kolchinsky2017} we can write an upper and a lower bound starting from the estimator
\begin{equation}
\label{eqn:entropy_estimator}
    \hat{H}_\mu = \sum_i \pi_i H(P_{\mu i}^s) - \sum_i \pi_i \log \left[\sum_{j}\pi_j e^{-d(P_{\mu i}^s || P_{\mu j}^s)}\right]
\end{equation}
where $d(P_{\mu i}^s || P_{\mu j}^s)$ is any distance function in the probability distributions space. We note that
\begin{align*}
    H\left(\prod_{\mu = 1}^N P_{\mu i}^s\right) & = -\int dx_1 \dots dx_N \, \prod_{\mu = 1}^NP_{\mu i}^s(x_\mu) \log \left[\prod_{\mu = 1}^NP_{\mu i}^s(x_\mu)\right] = \sum_{\mu = 1}^N H(P_{\mu i}^s)
\end{align*}
so the first part of Eq. (\ref{eqn:entropy_estimator}) for the entropy of the joint probability distribution is exactly equal to the sum of the estimators of the entropy of the one variable distributions. Thus, in the corresponding estimator for the multivariate information we are left with
\begin{align}
\label{eqn:estimator_multivariate}
    \hat{I}_{N,\rm{env}} = - \sum_i \pi_i \log \frac{\prod_{\mu = 1}^N\left(\sum_{j}\pi_j e^{-d_1(P_{\mu i}^s || P_{\mu j}^s)}\right)}{\sum_{j}\pi_j e^{-d_2(\prod_{\mu = 1}^N P_{\mu i}^s || \prod_{\mu = 1}^N P_{\mu j}^s )}}
\end{align}
where we denote as $d_1(\cdot ||\cdot)$ the distance function we choose for the one variable entropies and as $d_2(\cdot ||\cdot)$ the distance function we choose for the $N$ variables entropy.

Following \cite{Kolchinsky2017}, a lower bound for the entropy is achieved when we choose as a distance function the Chernoff-$\alpha$ divergence
\begin{equation*}
    C_\alpha(p || q) = -\log \int dx \,  p^\alpha(x) q^{1-\alpha}(x)
\end{equation*}
for any $\alpha \in [0,1]$, and an upper bound is instead achieved when we use a simple Kullback-Leibler divergence
\begin{equation*}
    D_{KL}(p || q) = \int dx\, p(x) \log\frac{p(x)}{q(x)}.
\end{equation*}
Therefore, Eq. \ref{eqn:estimator_multivariate} is a lower bound if we choose the Chernoff-$\alpha$ divergence for the one variable entropy and the Kullback-Leibler divergence for the $N$ variables entropy, and it is an upper bound is we make the opposite choice.

Both this upper and lower bound saturate in two particular cases. The first is the one in which $C_\alpha(\cdot || \cdot)$ diverges for all $i \ne j$. In fact, the Jensen inequality implies that $C_\alpha(\cdot || \cdot) \le (1-\alpha) D_{KL} (\cdot || \cdot)$, hence if the Chernoff-$\alpha$ divergence diverges so does the Kullback-Leibler divergence. In this case, the estimator of the mutual information is exact and we find
\begin{align}
    \label{eqn:residual_mutual}
    I_{N,\rm{env}}^{\infty} = -\sum \pi_i \log \frac{(\pi_i)^N}{\pi_i} = (N-1) H_\text{jumps}.
\end{align}
Qualitatively, this means that the probability distribution $P_{\mu i}^s(x)$ is infinitely different from $P_{\mu j}^s(x)$, so the discrete jumps between the $D_i$ states generate an infinitely different dynamics in terms of its stationary states.

The second, albeit trivial, case is the one in which the distances between both $P_{\mu i}^s$ and $P_{\mu j}^s$ are zero. Once more, both the upper and the lower bounds given by Eq. (\ref{eqn:estimator_multivariate}) saturate and we find
\begin{align}
    \label{eqn:residual_mutual_zero}
    I_{N,\rm{env}}^0 = -\sum \pi_i \log 1 = 0
\end{align}
which amounts to the trivial statement that if the two mixtures of Eqs. (\ref{eqn:mixture_1D}-\ref{eqn:mixture_2D}) have the same components they are also factorizable.

These results have a nice intuitive explanation. In fact, as long as $D_i$ is fixed the processes described by Equation (\ref{eqn:langevin_equations}) are independent and thus they cannot share any information. The only moment in time in which they are effectively coupled is when a jump $D_i \to D_j$ happens, when they share the sudden change in the diffusion coefficient - from then on, as long as $D_j$ is fixed, they evolve independently once more. As these changes are instantaneous, the greatest amount of information the processes can share corresponds to the entropy of the jumps, which is achieved when the processes are infinitely distinguishable for different diffusion coefficients $D_i$. In terms of information theory, the entropy of the jumps corresponds to our ignorance of the system, that is, since the jumps are stochastic we do not know when they happen.

\section*{D. Bounds on the mutual information for two non-interacting Ornstein-Uhlenbeck processes}
We now focus on the model proposed in the main text, in the non-interacting case. The stationary solution in the limit of slow-jumps is the Gaussian mixture
\begin{align}
\label{eqn:joint_OU_process}
    p_{12}(x_1, x_2) = \frac{1}{2\pi \tau}\left[\frac{\pi_-}{D_-} e^{-\frac{1}{2\tau D_-}\left(x_1^2 + x_2^2\right)} + \frac{\pi_+}{D_+} e^{-\frac{1}{2\tau D_+}\left(x_1^2 + x_2^2\right)}\right] = \pi_- \mathcal{N}(0, \Sigma_-) + \pi_+\mathcal{N}(0, \Sigma_+)
\end{align}
where $\pi_{+(-)}^s = w_{+(-)}/(w_+ + w_-)$ and $\Sigma_{-(+)} = D_{-(+)}\text{diag}\left(\tau, \tau\right)$. Similarly,
\begin{align}
\label{eqn:single_OU_process}
    p_1(x_1) = \frac{1}{\sqrt{2\pi\tau}}\left[\frac{\pi_-}{\sqrt{D_-}} e^{-\frac{x_1^2}{2\tau D_-}} + \frac{\pi_+}{\sqrt{D_+}} e^{-\frac{x_1^2}{2\tau D_+}}\right] = \pi_- \mathcal{N}(0, \tau D_-) + \pi_+\mathcal{N}(0, \tau D_+).
\end{align}
We are now interested in both the Chernoff-$\alpha$ divergence and the Kullback-Leibler divergence between the components of these Gaussian mixtures. In particular, for the two one-dimensional components of Eq. (\ref{eqn:single_OU_process}) we have
\begin{equation*}
    C_\alpha(\mathcal{N}(0, D_+) || \mathcal{N}(0, D_-)) = \frac{1}{2}\log\frac{(1-\alpha) + \alpha (D_-/D_+)}{(D_-/D_+)^\alpha}
\end{equation*}
which depends only on the ratio $D_-/D_+ := \varepsilon$. Since we are free to choose $\alpha$, we take $\partial_\alpha C_\alpha = 0$ so that the Chernoff divergence is minimum. We find
\begin{align*}
    \alpha = \frac{1- \varepsilon - \varepsilon\log\varepsilon}{(\varepsilon-1)\log\varepsilon} \implies C(a || b) = \frac{1}{2}\left[-1 + \log \frac{(\varepsilon-1) \varepsilon^\frac{1}{\varepsilon -1}}{\log \varepsilon}\right]:= \frac{1}{2}\,  z(\varepsilon).
\end{align*}
Similarly, for the components of Eq. \ref{eqn:joint_OU_process} we have 
\begin{equation*}
    C_\alpha(\mathcal{N}(0, \Sigma_+) || \mathcal{N}(0, \Sigma_-)) = \log\frac{(1-\alpha) + \alpha (D_+/D_-)}{(D_+/D_-)^\alpha}
\end{equation*}
and upon optimization over $\alpha$ we find the same result as before, up to a factor $1/2$, namely
\begin{equation*}
    C(\mathcal{N}(0, \Sigma_+) || \mathcal{N}(0, \Sigma_-))) = -1 + \log \frac{(\varepsilon-1) \varepsilon^\frac{1}{\varepsilon -1}}{\log \varepsilon} = z(\varepsilon).
\end{equation*}
Notice that $z(\varepsilon) = z(1/\varepsilon)$ implies that the distance between the component with a diffusion coefficient $D_-$ and the one with a diffusion coefficient $D_+$ is the same as the reversed one. We also note that the function $z(\varepsilon)$ has the following properties:
\begin{gather*}
    \lim_{\varepsilon \to 0} z(\varepsilon) = +\infty = \lim_{\varepsilon \to 0} z(1/\varepsilon) \\
    \lim_{\varepsilon \to 1} z(\varepsilon) = 0
\end{gather*}
which means that if $D_- \ll D_+$ the Chernoff divergence between the components of the Gaussian mixtures diverge.

We also need to write down explicitly the Kullback-Leibler divergences between the mixture components, which are
\begin{align*}
    D_{KL}(\mathcal{N}(0, D_+) || \mathcal{N}(0, D_-)) = \frac{1}{2}\left[\frac{1-\varepsilon}{\varepsilon}+\log\varepsilon\right] := \frac{1}{2} \, h(\varepsilon)
\end{align*}
and
\begin{align*}
    D_{KL}(\mathcal{N}(0, \Sigma_+) || \mathcal{N}(0, \Sigma_-)) = \frac{1-\varepsilon}{\varepsilon}+\log\varepsilon = h(\varepsilon).
\end{align*}
These distances are not symmetric anymore, but the function $h(\varepsilon)$ is such that
\begin{gather*}
    \lim_{\varepsilon \to 0} h(\varepsilon) = +\infty = \lim_{\varepsilon \to 0} h(1/\varepsilon) \\
    \lim_{\varepsilon \to 1} h(\varepsilon) = 0
\end{gather*}
so the limit $D_- \ll D_+$ is, perhaps unsurprisingly, the limit in which the distances between the mixture components diverge and the mutual information is exactly equal to the jump entropy.

Overall, the bounds on the mutual information are given by
\begin{align}
\label{eqn:OU_bounds}
    I_{\rm env}^{S,\rm{up}}\left(\frac{D_-}{D_+}, \frac{w_-}{w_+}\right) = -\pi_+\log\frac{\left[\pi_+ + \pi_- e^{-\frac{h(D_-/D_+)}{2}}\right]^2}{\pi_++\pi_- e^{-z(D_-/D_+)}} -\pi_-\log\frac{\left[\pi_+ e^{-\frac{h(D_+/D_-)}{2}} + \pi_-\right]^2}{\pi_+ e^{-z(D_+/D_-)} + \pi_-} \nonumber \\
    I_{\rm env}^{S,\rm{low}}\left(\frac{D_-}{D_+}, \frac{w_-}{w_+}\right) = -\pi_+\log\frac{\left[\pi_+ + \pi_- e^{-\frac{z(D_-/D_+)}{2}}\right]^2}{\pi_++\pi_- e^{-h(D_-/D_+)}} -\pi_-\log\frac{\left[\pi_+ e^{-\frac{z(D_+/D_-)}{2}} + \pi_-\right]^2}{\pi_+ e^{-h(D_+/D_-)} + \pi_-}
\end{align}
and they only depend on the ratios $D_-/D_+$ and $w_+/w_-$.

\section*{E. Disentangling the environment and the internal interactions in the mutual information}
Let us now consider the corresponding generalization to $N$ variables and $M$ jumps of the interacting case studied in the main text,
\begin{align}
    \dv{x_\mu}{t} = -\sum_{\nu}A_{\mu\nu} x_\nu + \sqrt{2 D_{i(t)}} \xi_\mu(t)
\end{align}
for $\mu = 1, \dots, N$ and $i  = 1, \dots, M$. In the slow-jumps limit, the mixture components are multivariate Gaussian distributions with a covariance matrix $\bm{\Sigma}$ that solves the Lyapunov equation
\begin{equation}
    \bm{A}\bm{\Sigma}_i + \bm\Sigma_i \bm{A}^T = 2D_i \mathbb{1}
\end{equation}
which we can rewrite as in the main text as
\begin{equation}
\label{eqn:lyapunov_equation_S}
    \bm{A}\tilde{\bm\Sigma} + \tilde{\bm\Sigma} \bm{A}^T = 2\tau^{-1} \mathbb{1}
\end{equation}
where $\bm\Sigma_i = D_i \tau \tilde{\bm\Sigma}$. Thus, the covariance matrix receives separate contributions from the diffusion coefficient $D_i$ and the interactions $\bm A$.

In order to compute the bounds,  we need the divergences
\begin{equation}
    C_\alpha(\bm\Sigma_i || \bm \Sigma_j) = \frac{1}{2}\log\frac{\det \left[(1-\alpha)\bm\Sigma_i+ \alpha\bm\Sigma_j)\right]}{\det^{1-\alpha} \bm\Sigma_i \det^{\alpha} \bm\Sigma_j} = \frac{1}{2}\log\frac{\det \tau \tilde{\bm\Sigma}\left[(1-\alpha)D_i+ \alpha D_j\right]}{\det^{1-\alpha} \tau D_i\tilde{\bm\Sigma} \det^{\alpha} \tau D_j\tilde{\bm\Sigma}} = \frac{N}{2}\log \frac{\left[(1-\alpha)D_i+ \alpha D_j\right]}{D_i^{1-\alpha}D_j^\alpha}
\end{equation}
and
\begin{align}
    D_{KL}(\bm\Sigma_i || \bm \Sigma_j) & = \frac{1}{2}\left[\log\frac{\det\bm\Sigma_j}{\det\bm\Sigma_i} + \Tr\bm\Sigma_j^{-1}\bm\Sigma_i - N\right] =  \frac{1}{2}\left[\log\frac{\det\tau D_j\tilde{\bm\Sigma}}{\det\tau D_i\tilde{\bm\Sigma}} + \Tr \frac{1}{\tau D_j}\tilde{\bm\Sigma}^{-1}\tau D_i\tilde{\bm\Sigma} - N\right] = \nonumber \\
    & = \frac{N}{2}\left[\log\frac{D_j}{D_i} + \frac{D_j}{D_i} - 1\right].
\end{align}
If we set $N=2$ we recover the two variables case considered in the main text, but in general these results hold for any $N$. As we can see, due to the factorization of the covariance matrix the bounds are the same as the ones of the non interacting case and they only depend on the ratios $D_i/D_j$.

Then, if we want to compute the multivariate information $I^{(N)}$ we need the entropies of the mixture components
\begin{equation}
    H^{(i)}_{1, \dots, N} = \frac{1}{2}\left[N \log(2\pi e \tau D_i) + \log\det\tilde{\bm\Sigma} \right]
\end{equation}
and
\begin{equation}
    H^{(i)}_\mu = \frac{1}{2}\left[\log(2\pi e \tau D_i) + \log\tilde{\bm\Sigma}_{\mu\mu} \right].
\end{equation}
Due to the interactions, it is not anymore the case in which $H^{(i)}_{1, \dots, N}$ is exactly equal to $\sum_{\mu} H^{(i)}_\mu$ and thus the bounds on the mutual information become
\begin{align}
\label{eqn:bound_interacting}
    I^{S,\rm{up/low}}_N = \frac{1}{2} \log\left[\frac{\prod_\mu \tilde{\bm\Sigma}_{\mu\mu}}{\det \tilde{\bm\Sigma}}\right] + I_{N,\rm{env}}^{S,\rm{up/low}}\left(\left\{\frac{D_i}{D_j}\right\}, \left\{\pi_i\right\}\right)
\end{align}
so the contribution of the interactions is disentangled from the one of the switching environment. Hence, in the limit in which all the distances between the mixture components diverge, we are left with
\begin{equation}
 I_N \to \frac{1}{2} \log\left[\frac{\prod_\mu \tilde{\bm\Sigma}_{\mu\mu}}{\det \tilde{\bm\Sigma}}\right] - (N-1) \sum_{i \in \pm} \pi_i \log\pi_i.
\end{equation}

Finally, in the case studied in the main text the interaction matrix is given by
\begin{equation}\renewcommand*{\arraystretch}{1.2}
    \bm A = \begin{pmatrix}
                \frac{1}{\tau} & -g_1 \\
                -g_2 & \frac{1}{\tau}
            \end{pmatrix}
\end{equation}
hence the solution to the Lyapunov equation is the covariance matrix
\begin{equation}\renewcommand*{\arraystretch}{1.2}
    \tilde{\bm\Sigma} = \frac{1}{g_1 g_2 \tau^2-1}\begin{pmatrix}
                        \frac{1}{2}g_1 \tau^2(g_2-g_1) - 1 & -\frac{1}{2}\tau(g_1+g_2) \\
                        -\frac{1}{2}\tau(g_1+g_2) & \frac{1}{2}g_2 \tau^2(g_1-g_2) - 1
                        \end{pmatrix}.
\end{equation}
Therefore, the result presented in the main text is simply given by
\begin{equation}
    \frac{1}{2} \log\left[\frac{\tilde{\bm\Sigma}_{11} \tilde{\bm\Sigma}_{22}}{\det \tilde{\bm\Sigma}}\right] = \frac{1}{2} \log\left[1 - \frac{4}{4 + \tau^2 (g_1 - g_2)^2}+ \frac{1}{1 - g_1 g_2 \tau^2} \right].
\end{equation}

Notice that in the fast-jumps limit, once we solve the Lyapunov equation Eq.~\eqref{eqn:lyapunov_equation_S}, the stationary probability distribution is the multivariate Gaussian distribution $\mathcal{N}(0, \ev{D}_\pi \tau \tilde{\bm \Sigma})$ that only depends on the single effective diffusion coefficient $\ev{D}_\pi$. In this limit we can compute the mutual information exactly
\begin{align}
    I_\mathrm{fast} = \frac{1}{2} \log\left[1 - \frac{4}{4 + \tau^2 (g_1 - g_2)^2}+ \frac{1}{1 - g_1 g_2 \tau^2} \right] = I_{\rm int}(g_1, g_2)
\end{align}
thus in this limit the only - constant - contribution to the mutual information is the first term of Eq. (\ref{eqn:bound_interacting}).

\section*{F. Multivariate information in the fast-jumps limit with non-linear interactions}
In the presence of non linear interactions the Fokker-Planck equation of the system reads
\begin{align}
    \label{eqn:fokker_planck_non_linear}
    \partial_t p_i(\vb{x}, t) & = \mathcal{L}_{\rm FP}^{(i)}\,p_i(\vb x, t) + \sum_{j = 1}^M \left[W(j \to i)  p_j(\vb{x}, t) - W(i \to j)  p_i(\vb{x}, t)\right] \\
    & = \sum_{\mu = 1}^N\partial_\mu\left[f_\mu(\vb x)p_i(\vb{x}, t)\right] + D_i\sum_{\mu= 1}^N \,\partial_\mu^2 \, p_i(\vb{x}, t) + \sum_{j = 1}^M \left[W(j \to i)  p_j(\vb{x}, t) - W(i \to j)  p_i(\vb{x}, t)\right]
\end{align}
where $\mathcal{L}_{\rm FP}^{(i)}$ is the Fokker-Planck operator and now $f_\mu(\vb x)$ contains the interaction terms between $x_\mu$ and all the remaining variables $x_{\nu \ne \mu}$. If we follow the same time-scale separation limits as in section B, the zero-th order stationary solution in the fast-jumps limit now solves the equation
\begin{align}
    0 = \sum_{i} \mathcal{L}_{\rm FP}^{(i)}\,\left[\pi_i p(\vb x, t)\right] = \sum_{\mu = 1}^N\partial_\mu\left[f_\mu(\vb x)p(\vb{x}, t)\right] + \ev{D}_\pi\sum_{\mu= 1}^N \,\partial_\mu^2 \, p(\vb{x}, t)
\end{align}
where $\ev{D}_\pi = \sum_i \pi_i \tilde{D}_i$. Although this equation cannot be solved exactly, the solution is not factorizable, and thus the multivariate information is not zero unless $f_\mu(\vb x) = f_\mu(x_\mu)$ which corresponds to the non-interacting case.

Hence, a non-vanishing multivariate information is a distinctive signature of underlying interactions. Notably, the main difference with respect to the previous linearized case corresponds to the fact that in the linear case it is possible to show that the multivariate information depends only on the interaction matrix $\vb A$, whereas in the general non-linear case we cannot factor out the dependence on the environments through $\ev{D}_\pi$.
\end{document}